\documentclass[a4paper,11pt]{article}

\pdfoutput=1

\usepackage{jcappub}
\usepackage[T1]{fontenc}
\usepackage{lmodern}
\usepackage{graphicx}
\usepackage{amsmath,amssymb,amsfonts}
\usepackage{bm}
\usepackage{slashed}
\usepackage{booktabs}
\usepackage{textcomp}

\title{Polarization Transport and Polarized Images of a Rotating Braneworld Black Hole}

\author[a]{Wen-Hao Deng,}
\author[a,1]{Qing-Quan Jiang,\note{Corresponding author.}}
\author[b]{Kai Lin,}
\author[a]{Pei Wang}

\affiliation[a]{School of Physics and Astronomy, China West Normal University, Nanchong 637009, People's Republic of China}
\affiliation[b]{Universidade Federal de Campina Grande, Campina Grande, PB, Brazil}

\emailAdd{qqjiangphys@yeah.net}

\abstract{ We investigate polarized synchrotron emission from an optically thin accretion flow around a rotating braneworld black hole, with particular emphasis on how higher-dimensional gravity modifies photon propagation and polarization transport. We directly integrate the null-geodesic equations together with the parallel-transport equations for the polarization four-vector and, within a common emissivity and magnetic-field prescription, examine the effects of the tidal charge, black-hole spin, observer inclination, and prograde or retrograde orbital motion. We analyze the linearly polarized intensity, electric-vector position angle (EVPA), Stokes parameters, the joint $Q$--$U$ polarization-state space, and the cumulative net linear polarization degree, and we compare representative braneworld configurations with Kerr and Kerr--Newman spacetimes. Within the parameter range considered here, the tidal charge and spin do not generate a qualitatively new large-scale polarization morphology. Instead, they primarily reorganize the polarized intensity, orientation, and polarization-state distribution in the strong-field region near the shadow boundary and photon ring. The viewing inclination controls the dominant projected morphology, while high inclinations enhance projection and Doppler asymmetries and make the distinction between prograde and retrograde flows more pronounced. The Stokes components and their joint state space expose local bending, stretching, and trajectory concentration that are less apparent in intensity images. The cumulative net linear polarization degree decreases with increasing cumulative radius, and larger tidal charge or spin lowers its amplitude for the representative models considered. Taken together, these complementary polarization observables provide a potential route for separating spacetime, viewing-geometry, and accretion-flow effects and for constraining braneworld tidal charge with horizon-scale polarimetry.}

\begin{document}
\maketitle

\section{Introduction}
\label{sec:intro}

The Event Horizon Telescope (EHT) resolved the asymmetric ringlike emission and central brightness depression of M87$^*$ on event-horizon scales, establishing a direct observational window on the strong-gravity region around a black hole~\cite{EHT2019}. Subsequent linear-polarization images revealed an ordered EVPA pattern across the ring and enabled constraints on the near-horizon magnetic-field topology, plasma properties, and accretion state~\cite{EHT2021b,EHT2021s}. Black-hole imaging has therefore progressed from identifying shadows and photon rings toward using intensity, polarization, and their spatial organization to probe the underlying spacetime. This requires a simultaneous understanding of the emission mechanism, relativistic motion of the accreting plasma, and photon propagation through curved spacetime.

Total-intensity imaging remains a basic probe of strong-field geometry, but its discriminating power is intrinsically limited. The shadow scale, photon-ring structure, brightness asymmetry, and frequency-shift distribution are determined not only by the metric but also by the dynamics and emissivity of the accretion flow. Metric deformations in rotating hairy black holes can alter the shadow size, photon-ring width, and redshift distribution~\cite{Meng2025}; in hairy Reissner--Nordstr\"om spacetimes, charge, hair, and deformation parameters can partially compensate one another and generate observational degeneracies~\cite{Meng2024}. Quantum-corrected regular black holes can likewise modify the shadow radius, photon ring, and observed intensity~\cite{Zhang2024}, while complementary observables such as time delays may distinguish rotating regular black holes even when image-position differences fall below the available angular resolution~\cite{Guo2024a}. These examples illustrate that similar intensity images need not imply identical spacetimes, and that an observed intensity difference cannot in general be attributed uniquely to the gravitational sector. Independent observables with a different response to propagation geometry are therefore valuable for testing departures from Kerr.

Braneworld gravity provides a particularly useful setting because the departure from four-dimensional general relativity has a direct higher-dimensional interpretation. In the Randall--Sundrum scenario, Standard Model fields are confined to a four-dimensional brane whereas gravity can propagate through the higher-dimensional bulk~\cite{Randall1999a,Randall1999b}. Dadhich et al. showed that a Reissner--Nordstr\"om-like geometry can arise on the brane even in the absence of a four-dimensional electric charge, with the charge-like term interpreted as a tidal imprint of the fifth-dimensional gravitational field~\cite{Dadhich2000}. Aliev and G\"umr\"uk\c{c}\"uo\u{g}lu subsequently generalized this construction to rotation~\cite{Aliev2005}. Although the rotating braneworld metric is formally similar to Kerr--Newman, its tidal charge $q$ is not a Maxwell charge; it represents a nonlocal projection of higher-dimensional Weyl curvature onto the brane and, importantly, can take negative values. As a metric-deformation parameter, $q$ modifies the horizon, photon region, circular-orbit structure, and rotational geometry, and thereby affects light bending, frequency shifts, and accretion-flow images. Previous imaging calculations have shown that spin, tidal charge, and inclination jointly modify the inner shadow, image asymmetry, redshift, and multi-band brightness distribution~\cite{Deng2026}. The tidal charge can also influence Blandford--Znajek energy extraction, with positive and negative values producing qualitatively different changes in the output power~\cite{Yang2025}. Thus $q$ links higher-dimensional gravity directly to strong-field orbital and radiative phenomena.

Polarization is especially useful because it retains information that is absent from total intensity. While $I$ primarily traces emissivity after gravitational lensing, Doppler boosting, and gravitational redshift, linear polarization also carries information about the orientation and degree of ordering of the electric field. It can be characterized by the Stokes parameters, the linear polarization degree, and the electric-vector position angle (EVPA). Early relativistic transfer calculations already demonstrated that strong-field propagation can rotate the polarization plane and modify the polarization degree~\cite{Connors1980}. For synchrotron emission, the polarization state at the source is determined by the local magnetic field, emitter four-velocity, and photon direction. After emission, the polarization four-vector is parallel transported along the corresponding null geodesic, so the spacetime curvature affects its projection onto the observer's screen along the full propagation path. Strongly lensed rays associated with the photon ring and higher-order images are therefore natural probes of the background geometry. Analytic ring models likewise show that magnetic-field orientation, fluid motion, gravitational lensing, and polarization transport jointly determine the observed pattern~\cite{Narayan2021,Gelles2021}.

A growing body of work has demonstrated the diagnostic potential of black-hole polarimetry. Synchrotron radiation from vortically moving charged particles near a magnetized Kerr black hole can produce characteristic polarized photon-ring structures~\cite{Zhang2023}; frame dragging, lensing, and gravitational Faraday rotation can separate the physical turning point of the flow from turning points in the primary image and EVPA pattern~\cite{Wang2025}. In rotating non-Kerr geometries, polarized intensity, EVPA, and $Q$--$U$ structures respond differently to spacetime deformations, spin, and magnetic-field orientation~\cite{Qin2025}. Similarly, in perfect-fluid-dark-matter spacetimes, the environmental deformation need not create a qualitatively new large-scale polarization morphology, but can systematically reorganize polarized intensity and EVPA near the photon ring~\cite{Huang2026}. These studies suggest that signatures of modified gravity may appear more naturally as stable, local changes in polarization direction, brightness distribution, and polarization-state organization than as an entirely different macroscopic image.

Numerically, the key task is to evolve the photon trajectory and polarization state with the same affine parameter and then project the transported polarization onto the observer's screen. The hidden symmetry underlying the Walker--Penrose constant provides an efficient analytic tool in Kerr-like spacetimes~\cite{Walker1970}, but methods tied to special algebraic structures are difficult to generalize. Covariant magnetoionic theory provides a systematic framework for polarized radiative transfer in curved spacetime~\cite{Broderick2004}, and public codes such as \texttt{ipole} evolve polarization in a fully relativistic setting~\cite{MoscibrodzkaGammie2018}. A more direct alternative is to integrate the null-geodesic and polarization parallel-transport equations as a coupled first-order ordinary differential equation (ODE) system. Recent Kerr--Newman calculations have used this strategy to obtain photon trajectories and polarization states without explicitly constructing a Killing tensor, revealing localized charge-dependent EVPA distortions near the photon ring~\cite{Li2026}. Covariant tools designed for more general spacetimes provide a complementary route to combining gravitational propagation with plasma effects~\cite{Huang2024}.

A systematic polarization study of rotating braneworld black holes is still lacking. Existing work has established that the tidal charge changes strong-field orbits, intensity images, and magnetospheric energetics, but has not followed how the polarization four-vector responds to the same geometric deformation. Because the observed polarization depends simultaneously on the accretion flow, emissivity, magnetic field, and propagation geometry, varying these ingredients together with $q$ would obscure the origin of any difference. We therefore keep the local emission and magnetic-field prescriptions fixed and allow the tidal charge to enter only through the metric and its Levi--Civita connection. This isolates the causal sequence in which the tidal charge modifies the spacetime, the null geodesics and polarization basis evolve differently, and the observable polarization is redistributed. The formal similarity between the rotating braneworld and Kerr--Newman metrics does not remove the need for this calculation: their charge-like parameters have different physical origins, and negative tidal charge has no direct counterpart in an ordinary electrically charged black hole.

In this work we investigate polarized synchrotron emission from an optically thin accretion flow around a rotating braneworld black hole. We directly couple the photon null-geodesic equations to the parallel transport of the polarization four-vector, construct the initial polarization from the emitter four-velocity and local magnetic field, and project the transported state onto the distant image plane. Keeping the local emissivity and magnetic-field model fixed, we study the effects of the tidal charge $q$, spin $a$, observer inclination, and prograde versus retrograde motion, and we compare representative braneworld configurations with Kerr and Kerr--Newman spacetimes. We analyze the linearly polarized intensity $I_{\rm pol}$, EVPA, Stokes $Q$ and $U$, the joint $Q$--$U$ state space, and a cumulative net linear polarization degree. Our aim is not to assume a qualitatively distinct braneworld image in advance, but to identify local, systematic signatures of tidal charge that can be cross-checked across complementary polarization observables.

\section{Rotating braneworld black hole and polarization transport}
\label{sec:geometry}

We first summarize the rotating braneworld geometry and the equations used for photon and polarization transport. In Boyer--Lindquist coordinates, the metric can be written as~\cite{Aliev2005,Aliev2009}
\begin{align}
 ds^{2}={}&-\left(1-\frac{2Mr-q}{\Sigma}\right)dt^{2}\nonumber\\
 &-\frac{2a(2Mr-q)\sin^{2}\theta}{\Sigma}\,dt\,d\phi\nonumber\\
 &+\frac{\Sigma}{\Delta}\,dr^{2}+\Sigma\,d\theta^{2}\nonumber\\
 &+\left(r^{2}+a^{2}\right)\sin^{2}\theta\,d\phi^{2}\nonumber\\
 &+\frac{a^{2}(2Mr-q)\sin^{4}\theta}{\Sigma}\,d\phi^{2}.
\label{eq:metric}
\end{align}
where
\begin{equation}
 \Delta=r^{2}-2Mr+a^{2}+q,
 \qquad
 \Sigma=r^{2}+a^{2}\cos^{2}\theta .
\label{eq:metric_functions}
\end{equation}
Here $M$ is the black-hole mass and $a=J/M$ is the spin parameter, with $J$ the angular momentum. The parameter $q$ is the tidal charge associated with nonlocal gravitational effects from the higher-dimensional bulk. It should not be interpreted as an electric charge confined to the brane. In contrast with the Kerr--Newman contribution $Q^{2}$, the tidal charge is not restricted to positive values, and negative values are allowed in braneworld models~\cite{Hou2021,Aliev2005,Aliev2009}. The metric is obtained under a Kerr--Schild ansatz for the induced four-dimensional brane geometry; whether it admits an exact embedding into a complete higher-dimensional bulk solution remains an open question. For $q=0$ the metric reduces to Kerr, while the nonrotating limit $a=0$ yields the corresponding static braneworld solution~\cite{Chamblin2000}.

For stationary and axisymmetric motion, the covariant momentum components $p_t=-E$ and $p_\phi=L$ are conserved. The photon wave vector is defined by
\begin{equation}
 k^{\mu}=\frac{dx^{\mu}}{d\lambda}
 =\left(\frac{dt}{d\lambda},\frac{dr}{d\lambda},\frac{d\theta}{d\lambda},\frac{d\phi}{d\lambda}\right),
\label{eq:kdef}
\end{equation}
where $\lambda$ is an affine parameter. The null geodesics satisfy
\begin{equation}
 \frac{dx^{\mu}}{d\lambda}=k^{\mu},
 \qquad
 \frac{dk^{\mu}}{d\lambda}=-\Gamma^{\mu}_{\ \nu\rho}k^{\nu}k^{\rho},
\label{eq:geodesic}
\end{equation}
with
\begin{equation}
 \Gamma^{\alpha}_{\ \mu\nu}=\frac{1}{2}g^{\alpha s}
 \left(\partial_{\mu}g_{s\nu}+\partial_{\nu}g_{s\mu}-\partial_s g_{\mu\nu}\right).
\end{equation}
The polarization four-vector $f^{\mu}$ is parallel transported along the same null geodesic,
\begin{equation}
 \frac{df^{\mu}}{d\lambda}=-\Gamma^{\mu}_{\ \nu\rho}k^{\nu}f^{\rho}.
\label{eq:parallel}
\end{equation}
Equations~\eqref{eq:geodesic} and \eqref{eq:parallel} form the basis of the numerical transport scheme. Using stationarity and axisymmetry to keep $p_t$ and $p_\phi$ fixed, we evolve six geodesic variables, $(t,r,\theta,\phi,p_r,p_\theta)$, together with the four components of $f^{\mu}$. The resulting reduced system therefore contains ten first-order ODEs. This is equivalent to the full geodesic system supplemented by the two conserved momenta and allows the photon orbit and polarization state to be evolved simultaneously.

In Kerr-type spacetimes, polarization transport can alternatively be treated with the Walker--Penrose constant when the required hidden symmetry is present~\cite{Narayan2021,Gelles2021}. Here we instead integrate Eq.~\eqref{eq:parallel} directly together with the null geodesics. Because the construction requires only the metric and its connection, the same numerical strategy can be extended to more general geometries without explicitly invoking a spacetime-specific polarization constant.

To maintain a continuous representation of the polarization direction in the numerical implementation, angular variables are mapped to fixed principal intervals. In particular, the azimuthal angle is taken modulo $2\pi$ and the linear-polarization direction modulo $\pi$, reflecting the invariance of a polarization axis under a $180^{\circ}$ reversal~\cite{Vincent2023}. This convention eliminates artificial discontinuities associated with branch choices in the angle representation.

We next construct a local orthonormal tetrad for the observer. With $\eta_{ab}=\mathrm{diag}(-1,1,1,1)$ and $g_{\mu\nu}e^{\mu}_{(a)}e^{\nu}_{(b)}=\eta_{ab}$, we use~\cite{Bardeen1972}
\begin{equation}
 e^{\mu}_{(0)}=\sqrt{-\frac{g_{\phi\phi}}{g_{tt}g_{\phi\phi}-g_{t\phi}^{2}}}
 \left(1,0,0,-\frac{g_{t\phi}}{g_{\phi\phi}}\right),
\label{eq:tetrad0}
\end{equation}
\begin{align}
 e^{\mu}_{(1)}&=\left(0,-\frac{1}{\sqrt{g_{rr}}},0,0\right), \notag\\
 e^{\mu}_{(2)}&=\left(0,0,\frac{1}{\sqrt{g_{\theta\theta}}},0\right), \notag\\
 e^{\mu}_{(3)}&=\left(0,0,0,-\frac{1}{\sqrt{g_{\phi\phi}}}\right).
\label{eq:tetradspatial}
\end{align}
These basis vectors are used to evaluate photon energies, redshift factors, and the observed polarization orientation. Following a fisheye-camera construction~\cite{Guo2024b,Vincent2011}, image-plane coordinates are mapped to a normalized field of view and then to a local propagation direction
\begin{equation}
 \bm{v}=(\cos\theta_x,\sin\theta_x\cos\psi_x,\sin\theta_x\sin\psi_x).
\label{eq:screen_direction}
\end{equation}
The corresponding initial photon momentum at the observer is written as
\begin{equation}
 k_{\rm obs}^{\mu}=-\kappa e_{(0)}^{\mu}+v^{(i)}e_{(i)}^{\mu},
\label{eq:initial_k}
\end{equation}
where $\kappa$ is fixed by the null condition $k_{\rm obs}^{\mu}k^{\rm obs}_{\mu}=0$.

\section{Emission model and initial polarization}
\label{sec:initial}

To complete the polarized ray-tracing calculation, we specify a phenomenological source function for the optically thin synchrotron emission. We adopt~\cite{Shcherbakov2010}
\begin{align}
 I_{\rm pol}(r,\theta,\nu)={}&\Pi_0\cos^{2}\theta
 \left[\frac{\nu^{3}}{\exp(\nu/T(r))-1}\right] \notag\\
 &\times\left[1+\alpha\tanh(\gamma x)\right]
 \frac{\exp\left[-\tfrac{1}{2}(\sinh^{-1}x)^{2}\right]}
 {(x^{2}+1)^{\delta/2}}, \notag\\
 x={}&\frac{r-\beta}{\sigma}.
\label{eq:source}
\end{align}
where $\beta$ sets the characteristic emission radius, $\sigma$ its radial width, and $\delta$ the outer radial falloff. The parameters $\alpha$ and $\gamma$ control the amplitude and radial scale of the asymmetric component. We use $\Pi_0$ for the source polarization fraction in order to distinguish it from the observed polarization degree defined below. The factor $\cos^{2}\theta$ describes the adopted angular dependence, $\nu$ is the photon frequency, and $T(r)$ is an effective radial temperature profile.

The local magnetic field determines the emitted polarization orientation. We use a parametrized helical field in the local frame,
$B^{\mu}(r)=(0,B^{r},B^{\theta},B^{\phi})$, with
\begin{align}
 B^{r}(r)&=\frac{b_{\rm ratio}}{r}, \notag\\
 B^{\theta}(r)&=\frac{{\rm amp}}{r}\sin(k\ln r), \notag\\
 B^{\phi}(r)&=\frac{1}{r}
 \left[1+\frac{1}{2}\cos(k\ln r)\right].
\label{eq:Bfield}
\end{align}
Here $b_{\rm ratio}$ controls the radial component, ${\rm amp}$ the polar component, and $k$ the radial oscillation scale of the helical structure. The local pitch angle may be characterized by
\begin{equation}
 \tan\psi(r)=\frac{\sqrt{[B^{r}(r)]^{2}+[B^{\theta}(r)]^{2}}}{B^{\phi}(r)}.
\label{eq:pitch}
\end{equation}
Large $\psi$ corresponds to a stronger nonazimuthal component, whereas small $\psi$ indicates a predominantly toroidal field.

The magnetic field in Eq.~\eqref{eq:Bfield} is a phenomenological field associated with the emitting accretion flow. It is physically distinct from the braneworld tidal charge. The latter is not a Maxwell charge and does not require an additional electromagnetic four-potential or an intrinsic black-hole magnetic field. In our setup, the magnetic prescription fixes the source-side polarization geometry, whereas $q$ acts through the spacetime metric and connection and therefore modifies the photon trajectory and the parallel transport of the polarization state.

Given the emission point $x^{\mu}$, photon momentum $k^{\mu}$, emitter four-velocity $U^{\mu}$, and local magnetic field $B^{\mu}$, the initial polarization four-vector is chosen transverse to the photon direction and normalized,
\begin{equation}
 k_{\mu}f^{\mu}=0,\qquad f_{\mu}f^{\mu}=1.
\label{eq:constraints}
\end{equation}
A covariant construction is
\begin{align}
 f^{\mu}&=\frac{\epsilon^{\mu\nu\rho\sigma}U_{\nu}k_{\rho}B_{\sigma}}
 {\mathcal N},
 \label{eq:initial_f}\\
 \mathcal N&=\Bigl|\bigl(\epsilon^{\alpha\beta\gamma\delta}
 U_{\beta}k_{\gamma}B_{\delta}\bigr)
 g_{\alpha\lambda}
 \bigl(\epsilon^{\lambda\eta\kappa\zeta}
 U_{\eta}k_{\kappa}B_{\zeta}\bigr)\Bigr|^{1/2}.
 \notag
\end{align}
At emission this choice is also orthogonal to the emitter four-velocity. The subsequent propagation is governed by Eq.~\eqref{eq:parallel}, which preserves transversality with respect to $k^{\mu}$ and the norm of $f^{\mu}$. In this way, the tidal charge affects the observed polarization only through the braneworld geometry, rather than through an artificial modification of the source magnetic field.

At the observer, the transported polarization vector is projected onto the two orthogonal screen directions,
\begin{equation}
 f^{(i)}=f^{\mu}e^{(i)}_{\mu},\qquad i=1,2.
\end{equation}
The linear Stokes parameters are
\begin{equation}
 Q=I_{\rm pol}\cos 2\chi,\qquad U=I_{\rm pol}\sin 2\chi,
\label{eq:QU}
\end{equation}
and the EVPA is
\begin{equation}
 \chi=\frac{1}{2}\operatorname{atan2}(U,Q).
\label{eq:evpa}
\end{equation}
Thus the joint solution of the null geodesics and polarization transport provides the observed $I$, $Q$, and $U$, from which the EVPA and polarization degree can be constructed.

\section{Ray tracing and polarized observables}
\label{sec:raytracing}

The total redshift factor between the emitter and distant observer is $g=\nu_{\rm obs}/\nu_{\rm em}$. Liouville's theorem implies the invariance of $I_{\nu}/\nu^{3}$ along a null geodesic, so the observed total and linearly polarized intensities are modeled as
\begin{equation}
 I=g^{3}j_I(r_{\rm em}),\qquad I_{\rm pol}=g^{3}j_p(r_{\rm em}),
\label{eq:intensity}
\end{equation}
where $j_I$ and $j_p$ are the local total and polarized emissivities. When the photon reaches the observer, the transported polarization four-vector is projected onto two orthogonal screen directions $e_{(x)}^{\mu}$ and $e_{(y)}^{\mu}$,
\begin{equation}
 E_x\propto-f_{\mu}e_{(x)}^{\mu},\qquad E_y\propto f_{\mu}e_{(y)}^{\mu}.
\end{equation}
The Stokes parameters can then be written as
\begin{align}
 Q&=I_{\rm pol}\frac{E_x^{2}-E_y^{2}}{E_x^{2}+E_y^{2}},\\
 U&=I_{\rm pol}\frac{2E_xE_y}{E_x^{2}+E_y^{2}}.
\label{eq:QU_screen}
\end{align}
The linear polarization degree is
\begin{equation}
 \Pi=\frac{\sqrt{Q^{2}+U^{2}}}{I}.
\label{eq:Pi}
\end{equation}

Each image pixel corresponds to an independent backward-traced null geodesic. The geodesic and polarization equations are integrated until the ray reaches the emitting region, approaches the horizon, exits the outer numerical boundary, or meets the prescribed geometric termination criterion. At the emission point the local intensity and polarization state are assigned, and the polarization vector is transported to the observer. We neglect circular polarization and focus on the linear Stokes parameters. The resulting rays are assembled into two-dimensional maps of total intensity, linearly polarized intensity, $Q$, $U$, EVPA, and polarization degree. Numerical stability is checked by increasing image resolution and tightening the integration tolerance, following the convergence strategy used in related polarized ray-tracing calculations~\cite{Li2026}.

For visualization, the intensity maps are normalized and passed through a smooth nonlinear mapping to compress their large dynamic range~\cite{EHT2019}. This display operation does not alter the relative Stokes information used to calculate EVPA or polarization degree. In all parameter comparisons below, the local emissivity and magnetic-field prescription are held fixed. We vary only the tidal charge $q$, spin $a$, observer inclination $\theta_0$, and the sense of orbital motion. This controlled setup is intended to isolate geometric propagation effects from changes in the source magnetic topology.

\subsection{Polarization vectors, EVPA, and synchrotron intensity}

\begin{figure}[htbp]
\centering
\includegraphics[width=.48\textwidth]{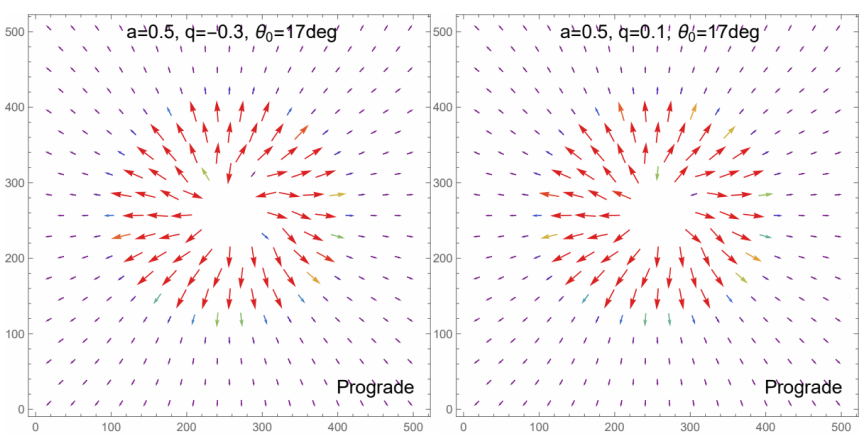}
\includegraphics[width=.48\textwidth]{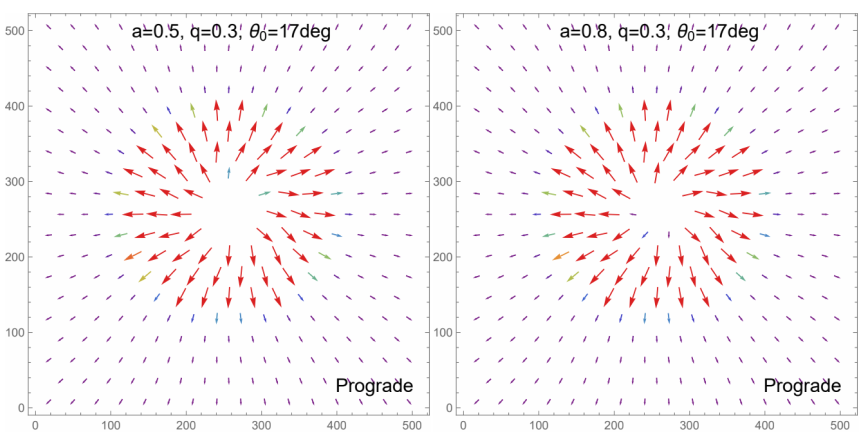}\\
\includegraphics[width=.48\textwidth]{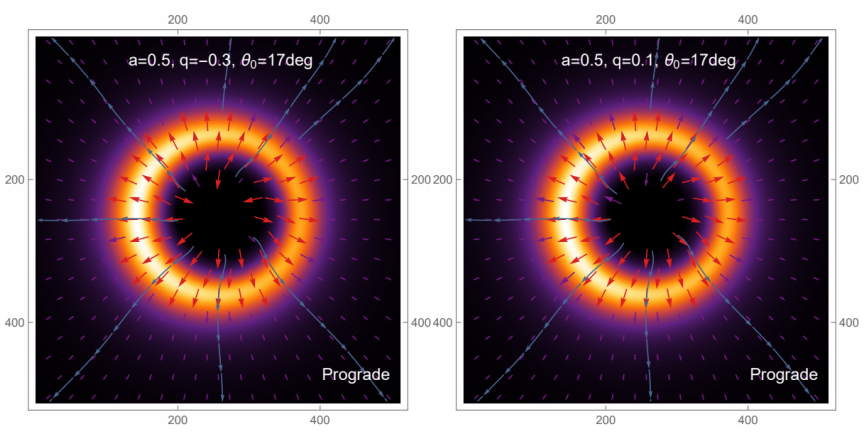}
\includegraphics[width=.48\textwidth]{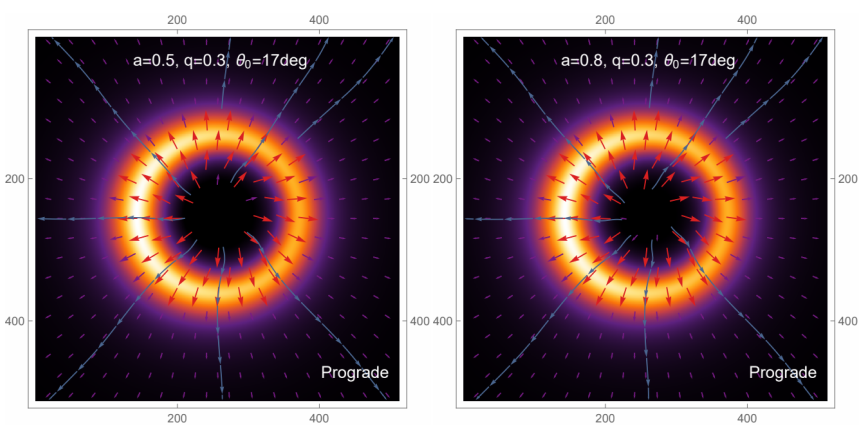}

\caption{Polarization vectors, EVPA, polarization streamlines, and synchrotron intensity for the prograde accretion flow around a rotating braneworld black hole with $M=1$ and $\theta_0=17^\circ$. From left to right, the columns correspond to $(a,q)=(0.5,-0.3)$, $(0.5,0.1)$, $(0.5,0.3)$, and $(0.8,0.3)$. The upper row shows the polarization-vector and EVPA distributions, while the lower row shows the polarization vectors and streamlines overlaid on the synchrotron-intensity images.}
\label{fig:1}
\end{figure}

\begin{figure}[htbp]
\centering
\includegraphics[width=.95\textwidth]{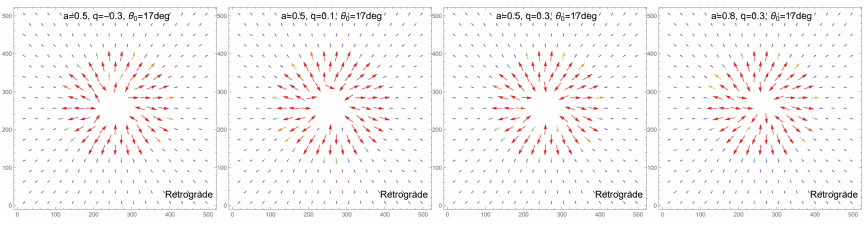}
\includegraphics[width=.95\textwidth]{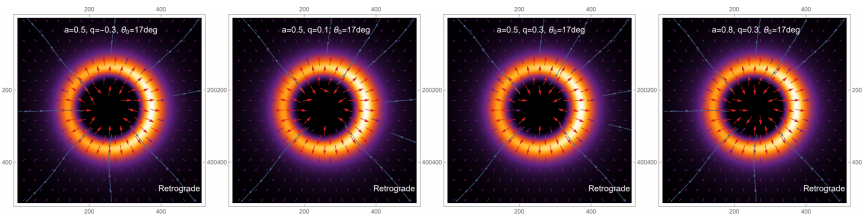}
\caption{Same as Fig.~\ref{fig:1}, but for the retrograde accretion flow.}
\label{fig:2}
\end{figure}

Figures~\ref{fig:1} and \ref{fig:2} show the synchrotron intensity, polarization vectors, polarization streamlines, and EVPA for prograde and retrograde flows at the fixed inclination $\theta_0=17^\circ$. A first robust feature is that varying either $q$ or $a$ within the range considered does not produce a qualitatively different large-scale polarization morphology. In both orbital configurations the emission retains a bright ring surrounding the central shadow, and the polarization vectors and streamlines preserve a continuous large-scale organization. The braneworld parameters therefore act mainly through local modifications of an existing polarization pattern rather than by reconstructing the entire polarization field.

For the prograde flow, increasing $q$ at fixed $a=0.5$ leaves the overall ring morphology nearly unchanged while shifting the local brightness and polarization orientation near the shadow boundary and bright ring. The directional fields become progressively more similar away from the central strong-field region. This behavior is consistent with the geometric role of the tidal charge: changing $q$ modifies the metric and Christoffel connection, which changes the null geodesics, lensing, propagation redshift, and the connection along which the polarization four-vector is parallel transported. The observed EVPA and streamline differences are therefore not caused by an additional electromagnetic field. They arise from the projection of higher-dimensional gravitational effects onto the four-dimensional brane geometry.

At fixed $q=0.3$, increasing the spin from $a=0.5$ to $0.8$ again preserves the overall morphology but produces additional local changes near the shadow. These differences are associated with the rotating geometry and frame dragging. It is important not to interpret this as the spin simply rotating the polarization vector by an extra angle. Rather, the spin changes the azimuthal photon motion and the full geometry along which the polarization state is transported, and the resulting difference is projected onto the observer's screen as a local rearrangement of EVPA and polarization streamlines.

The retrograde models show the same broad pattern: the large-scale structure remains similar while the local brightness and polarization orientation change near the black hole. The distinction between prograde and retrograde flows, however, has a different origin from changing $q$ or $a$. For fixed $(a,q)$ the background metric is identical, while the emitter four-velocity $U^{\mu}$ changes because the orbital angular momentum reverses relative to the black-hole spin. This modifies the local photon initial conditions, Doppler and total redshift factors, and the initial polarization constructed from $U^{\mu}$, $k^{\mu}$, and $B^{\mu}$. The rays then propagate through the same braneworld geometry. The prograde--retrograde difference therefore reflects the coupling between accretion-flow kinematics and the rotating spacetime rather than a change in the gravitational theory itself.

\subsection{Linearly polarized intensity}

\begin{figure}[htbp]
\centering
\includegraphics[width=.48\textwidth]{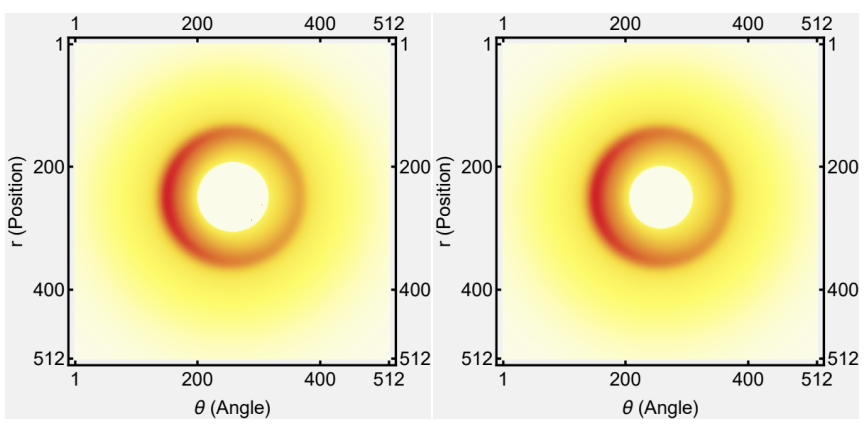}
\includegraphics[width=.48\textwidth]{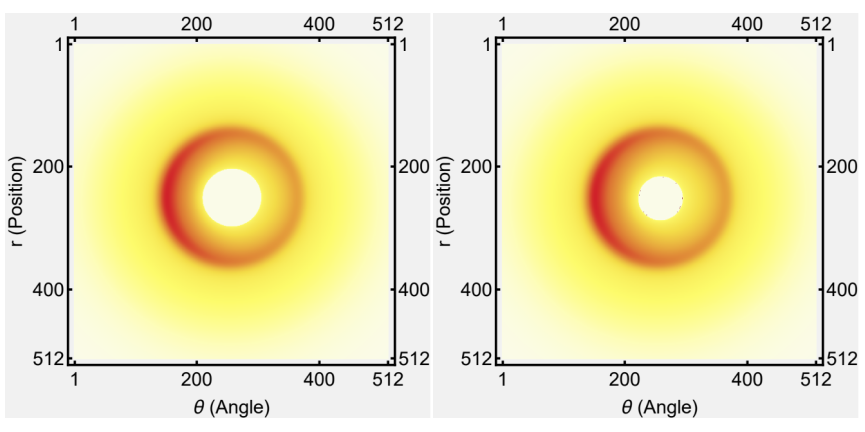}\\
\includegraphics[width=.24\textwidth]{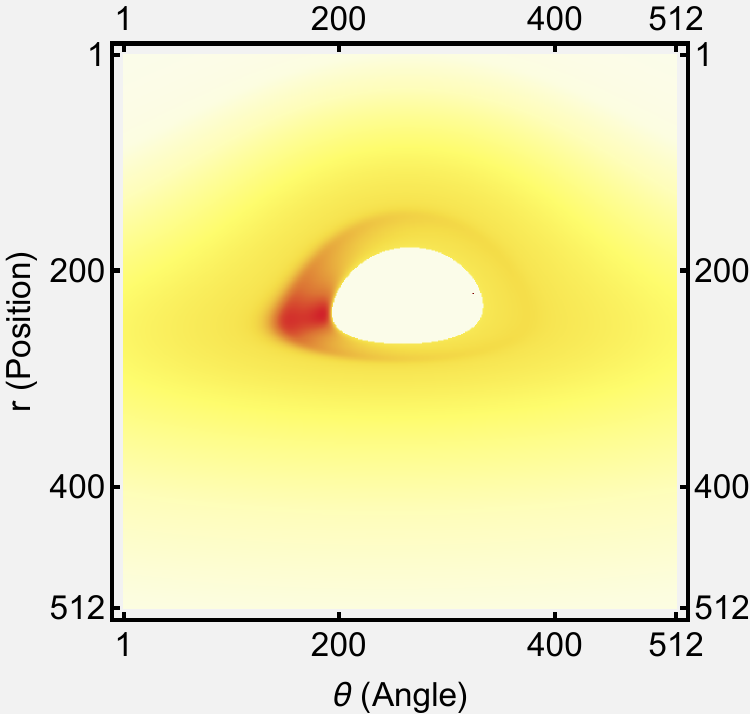}
\includegraphics[width=.24\textwidth]{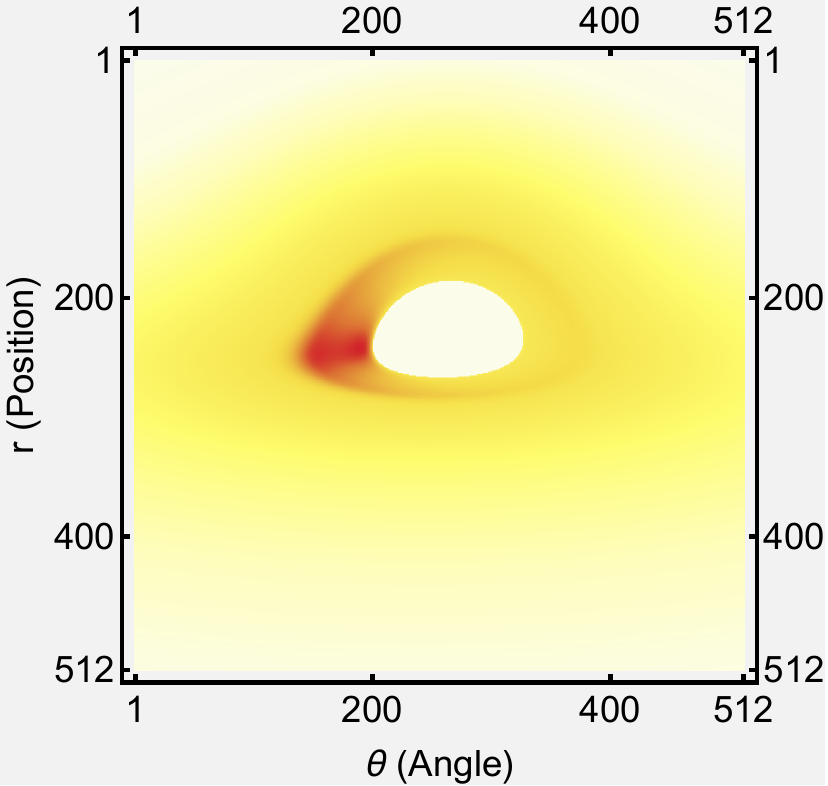}
\includegraphics[width=.24\textwidth]{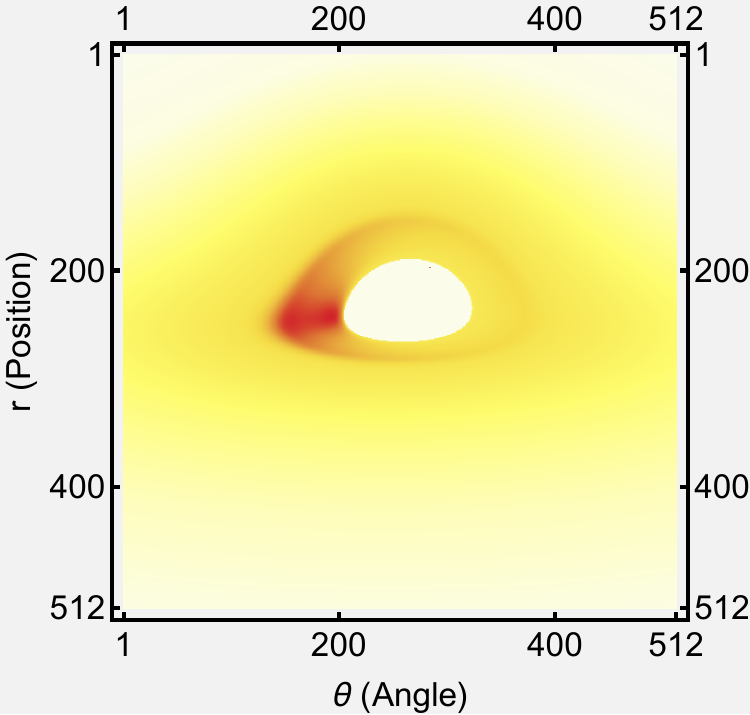}
\includegraphics[width=.24\textwidth]{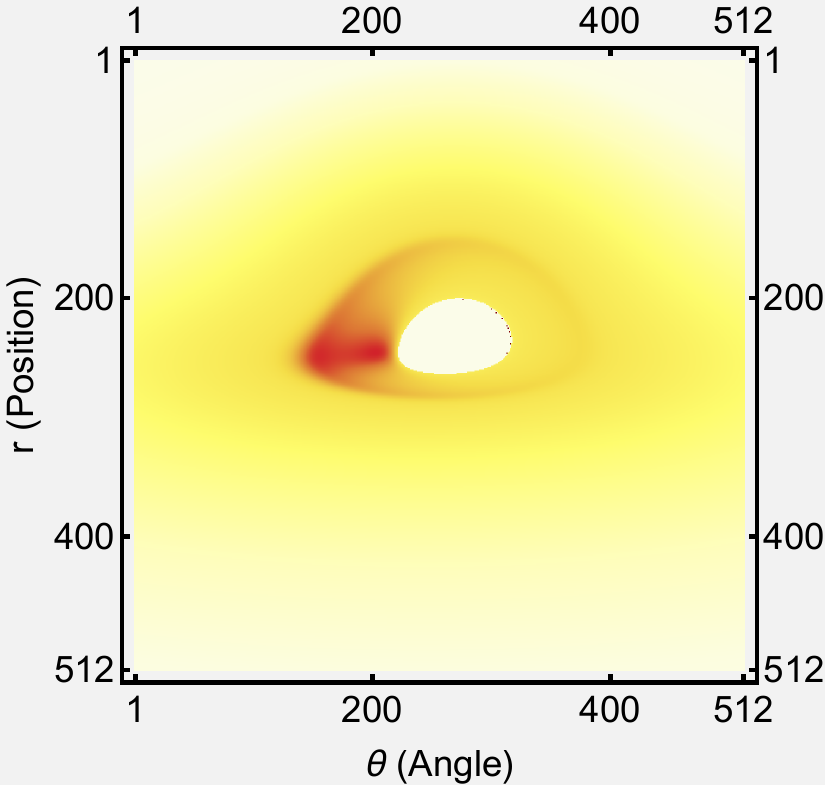}
\caption{Linearly polarized intensity for the prograde accretion flow around a rotating braneworld black hole with $M=1$. The upper and lower rows correspond to $\theta_0=17^\circ$ and $75^\circ$, respectively. From left to right, the columns show $(a,q)=(0.5,-0.3)$, $(0.5,0.1)$, $(0.5,0.3)$, and $(0.8,0.3)$. Variations in $q$ and $a$ mainly modify the local polarized-intensity distribution, whereas the inclination changes the global projected morphology.}
\label{fig:3}
\end{figure}

\begin{figure}[htbp]
\centering
\includegraphics[width=.95\textwidth]{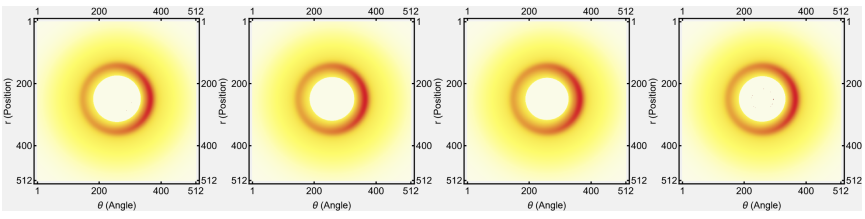}
\includegraphics[width=.95\textwidth]{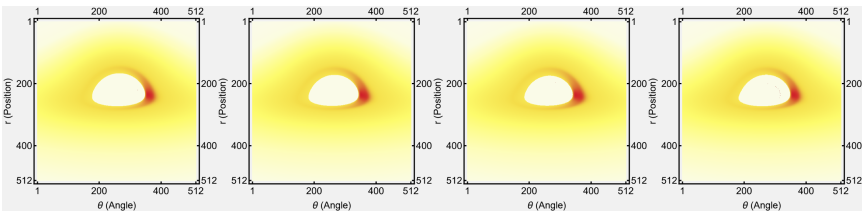}
\caption{Same as Fig.~\ref{fig:3}, but for the retrograde accretion flow.}
\label{fig:4}
\end{figure}

Figures~\ref{fig:3} and \ref{fig:4} show the linearly polarized intensity $I_{\rm pol}$ for prograde and retrograde flows. The most prominent trend is the strong dependence of the global morphology on observer inclination. At $\theta_0=17^\circ$, both orbital configurations retain a relatively regular, nearly ringlike polarized-emission pattern. At $\theta_0=75^\circ$, the image is strongly compressed and becomes a nonaxisymmetric arc or crescent with polarized emission concentrated toward one side. This change follows from the projected disk geometry, the larger line-of-sight component of the orbital velocity and hence stronger Doppler contrast, and a greater asymmetry among the strong-field propagation histories of rays from different parts of the disk.

At fixed inclination and $a=0.5$, changing $q$ preserves the overall ringlike or arclike structure but shifts the local polarized intensity near the shadow boundary and bright regions. Because the local emissivity and magnetic field are fixed, these differences primarily trace the $q$ dependence of the lens mapping and redshift factor. Increasing the spin at fixed $q=0.3$ likewise preserves the global structure while changing the location and amplitude of local bright features. We therefore find that the viewing inclination primarily controls the global projected morphology of $I_{\rm pol}$, while the tidal charge and spin mainly reorganize the polarized intensity in the strong-field region.

\subsection{Stokes $Q$ and $U$}

\begin{figure}[htbp]
\centering
\includegraphics[width=.48\textwidth]{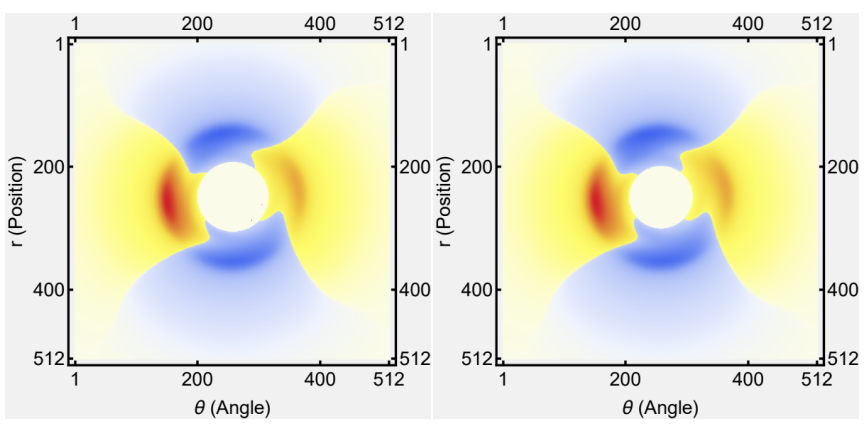}
\includegraphics[width=.48\textwidth]{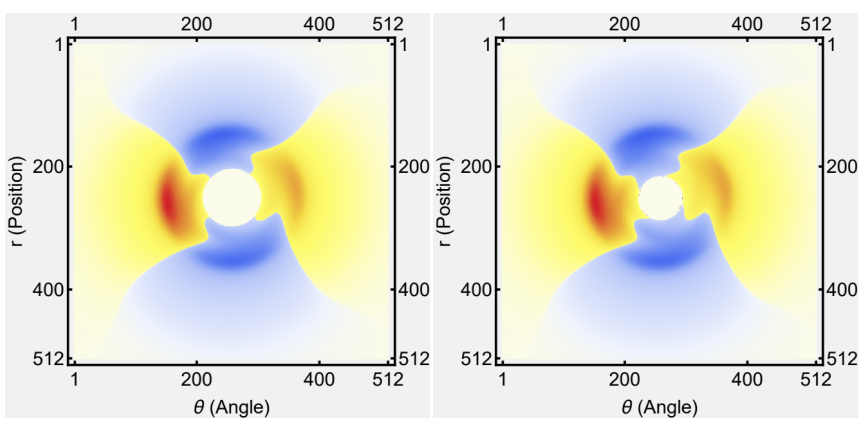}\\
\includegraphics[width=.24\textwidth]{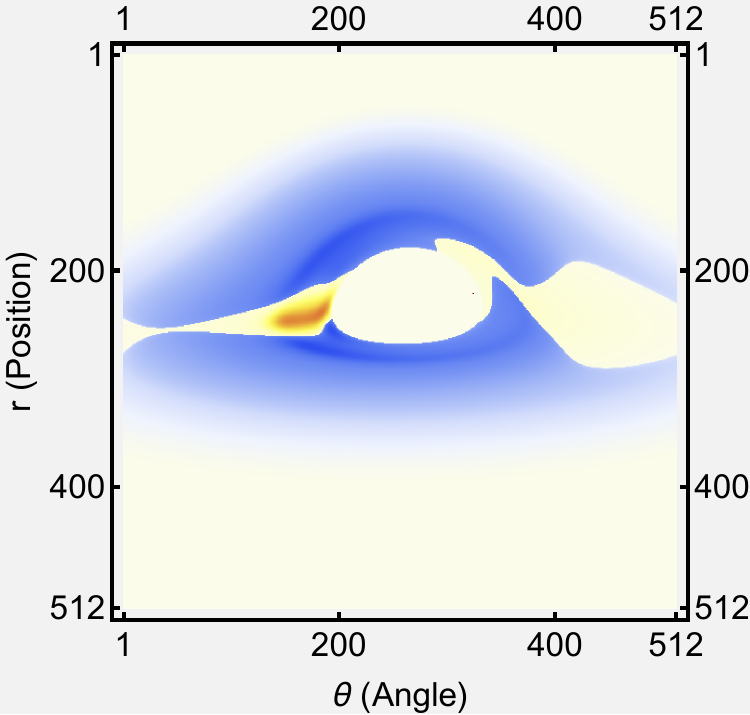}
\includegraphics[width=.24\textwidth]{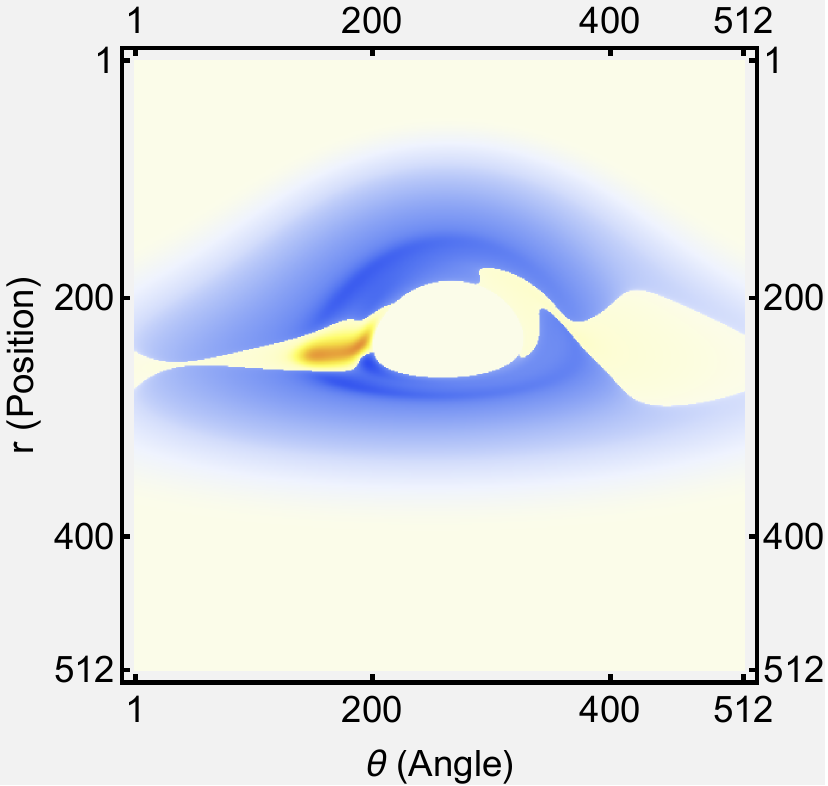}
\includegraphics[width=.24\textwidth]{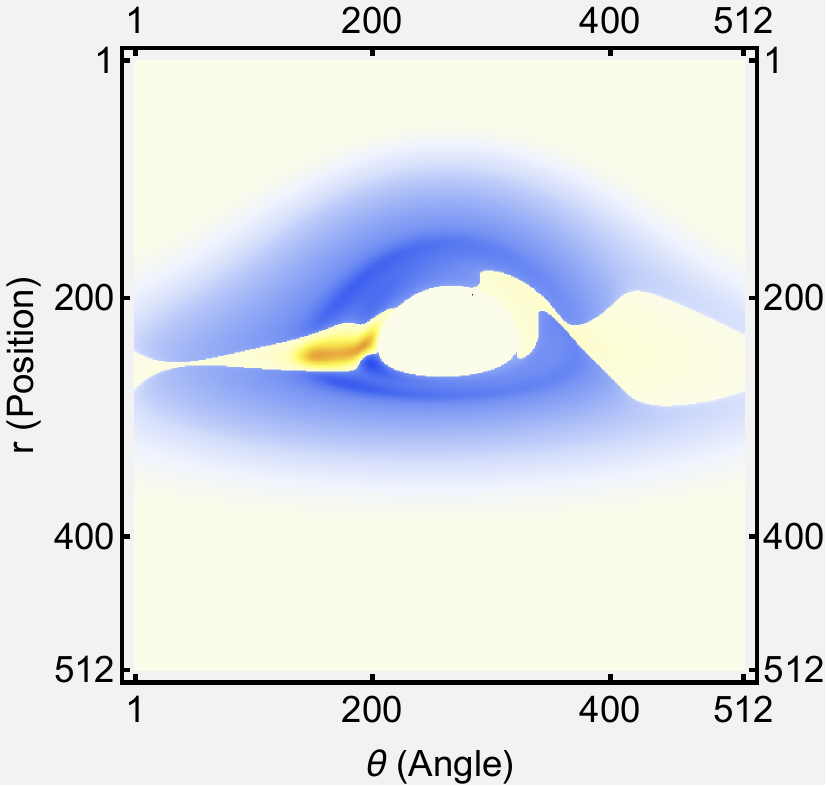}
\includegraphics[width=.24\textwidth]{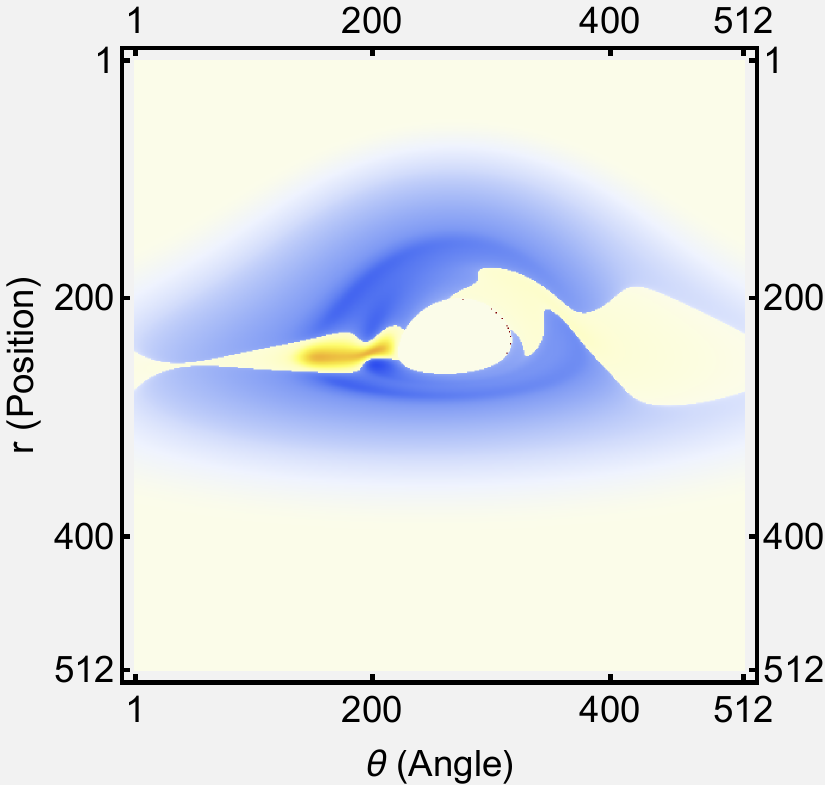}
\caption{Stokes $Q$ for the prograde accretion flow around a rotating braneworld black hole with $M=1$. The upper and lower rows correspond to $\theta_0=17^\circ$ and $75^\circ$, respectively. From left to right, the columns show $(a,q)=(0.5,-0.3)$, $(0.5,0.1)$, $(0.5,0.3)$, and $(0.8,0.3)$.}
\label{fig:5}
\end{figure}

\begin{figure}[htbp]
\centering
\includegraphics[width=.95\textwidth]{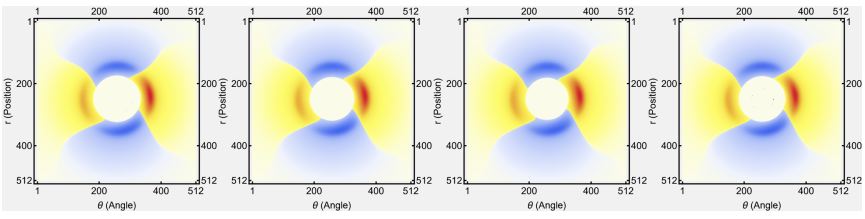}
\includegraphics[width=.95\textwidth]{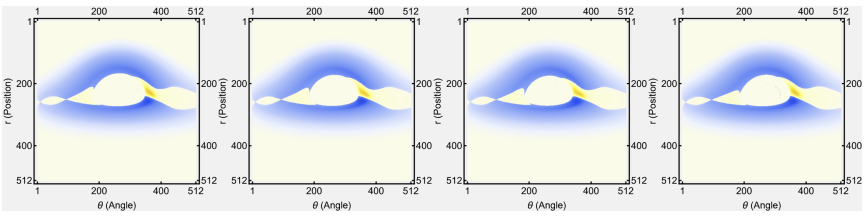}
\caption{Same as Fig.~\ref{fig:5}, but for the retrograde accretion flow.}
\label{fig:6}
\end{figure}

Although $I_{\rm pol}$ fixes the amplitude of the linearly polarized emission, it does not uniquely specify the polarization direction. Figures~\ref{fig:5} and \ref{fig:6} therefore show the Stokes-$Q$ maps. At the lower inclination, $Q$ exhibits a comparatively regular multiregion organization around the central shadow. At $\theta_0=75^\circ$, this structure is compressed toward the projected disk and becomes more horizontally layered. Changing $q$ at fixed $a=0.5$ leaves the global organization recognizable but shifts local enhancement regions, structural boundaries, and the width of the distribution near the black hole. Increasing $a$ produces analogous local adjustments in the central strong-field region.

These changes should not be attributed to either polarized intensity or EVPA alone. The tidal charge modifies both the local $I_{\rm pol}$ through lensing and redshift and the transported angle $\chi$ through the altered connection. The product $I_{\rm pol}\cos2\chi$ then maps these changes into the observed $Q$ pattern. The prograde and retrograde cases retain a similar large-scale organization but differ in the location and detailed shape of local structures, consistent with the different emitter kinematics in the same background geometry.

\begin{figure}[htbp]
\centering
\includegraphics[width=.48\textwidth]{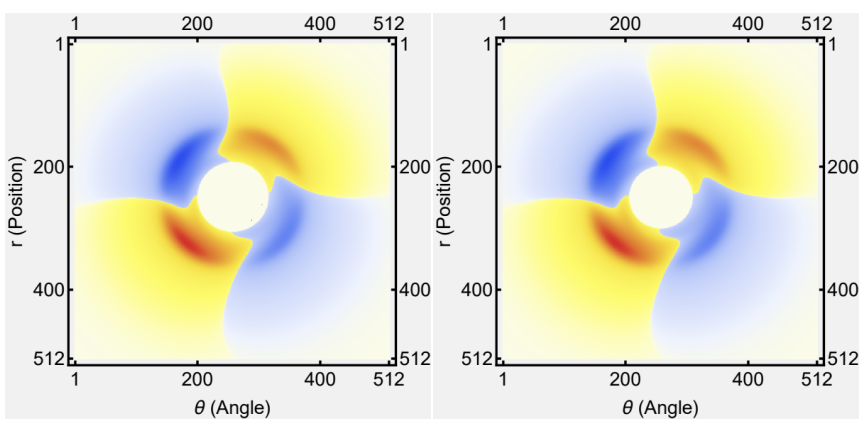}
\includegraphics[width=.48\textwidth]{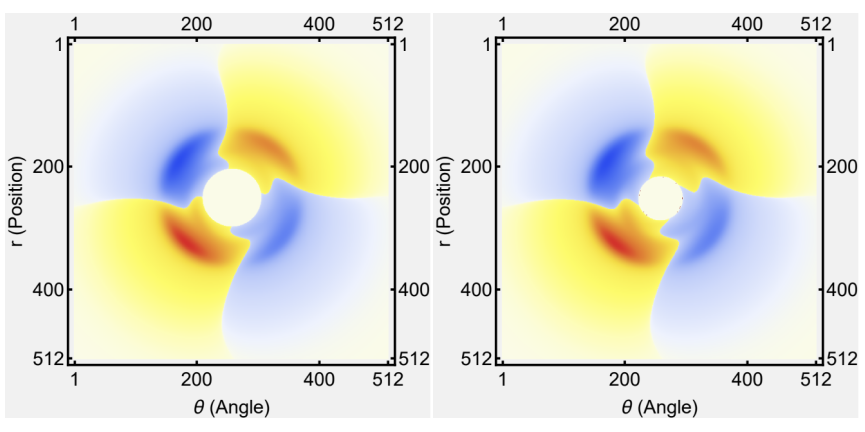}\\
\includegraphics[width=.24\textwidth]{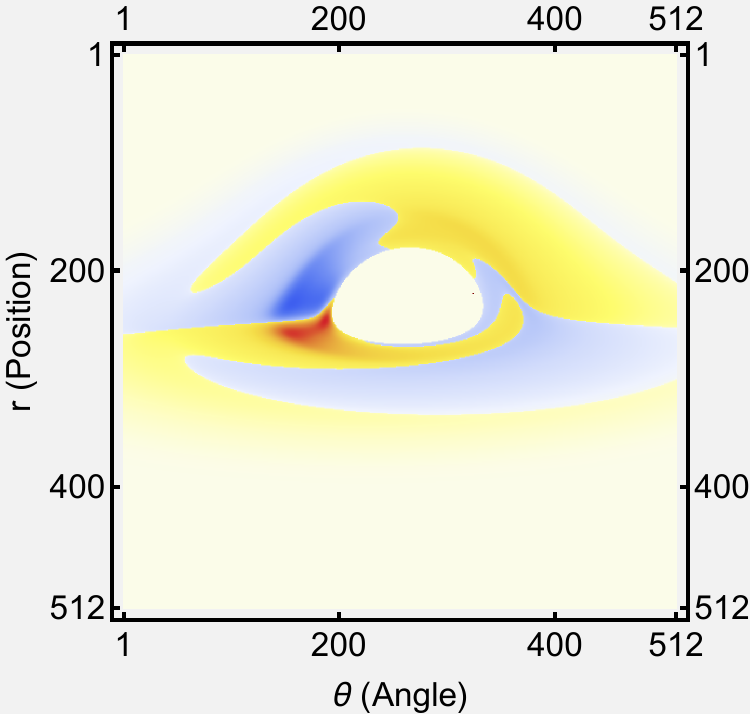}
\includegraphics[width=.24\textwidth]{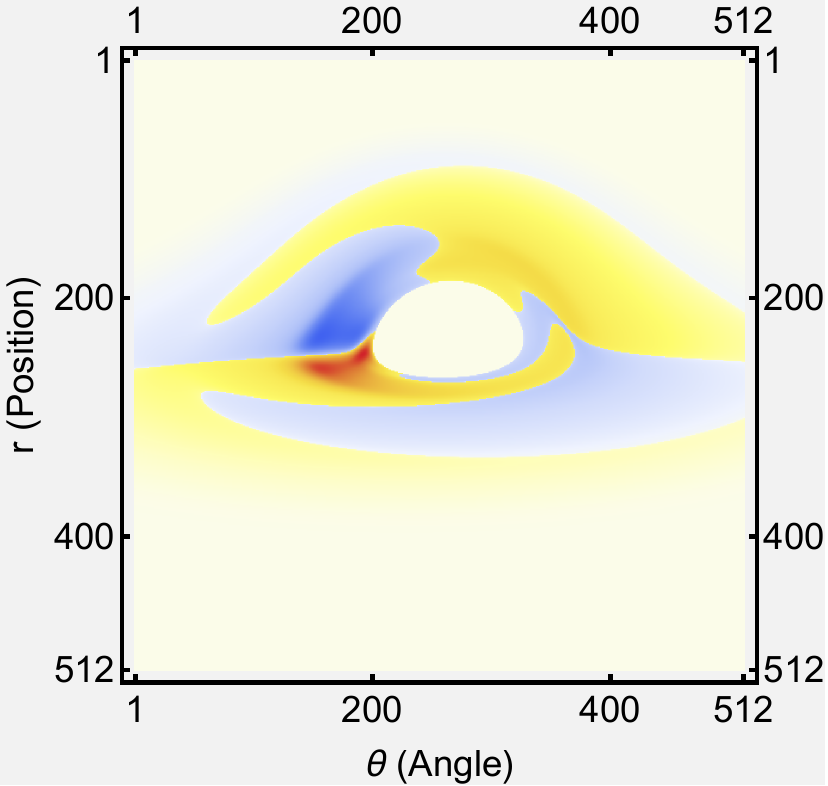}
\includegraphics[width=.24\textwidth]{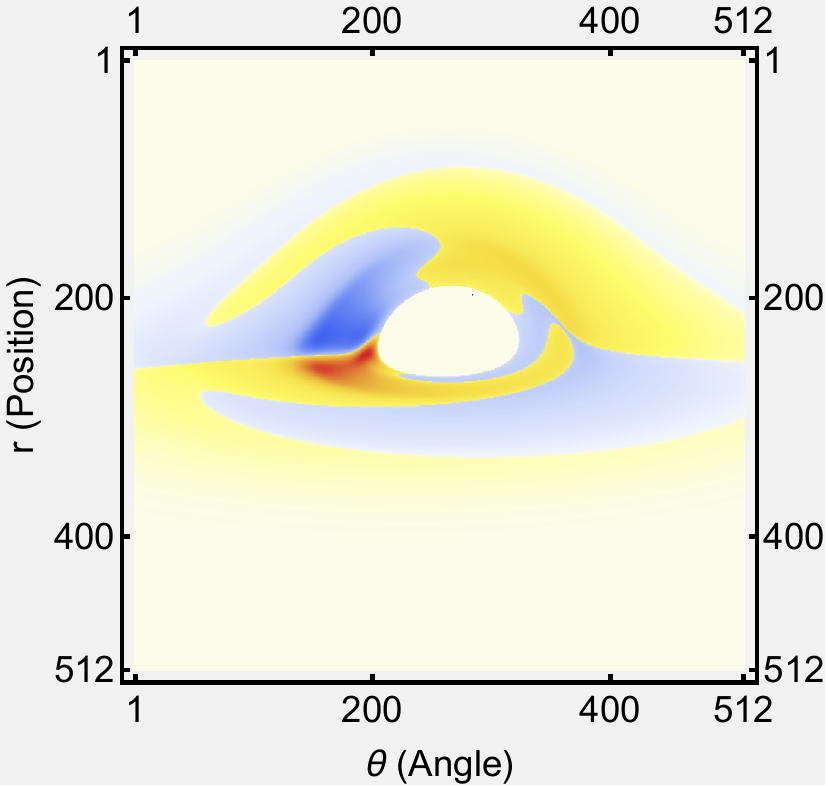}
\includegraphics[width=.24\textwidth]{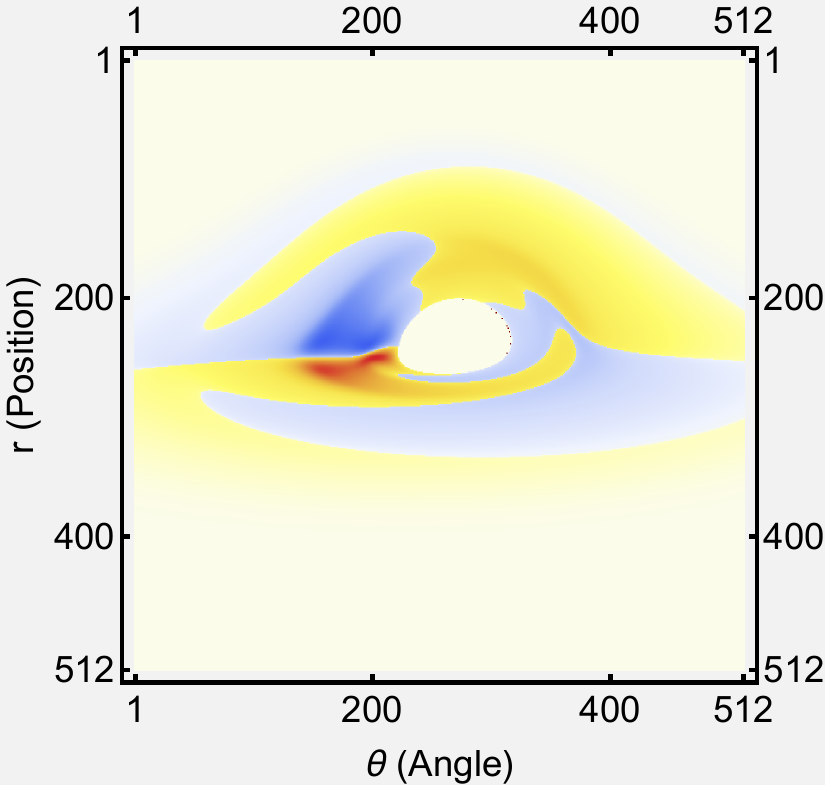}
\caption{Stokes $U$ for the prograde accretion flow around a rotating braneworld black hole with $M=1$. The upper and lower rows correspond to $\theta_0=17^\circ$ and $75^\circ$, respectively. From left to right, the columns show $(a,q)=(0.5,-0.3)$, $(0.5,0.1)$, $(0.5,0.3)$, and $(0.8,0.3)$.}
\label{fig:7}
\end{figure}

\begin{figure}[htbp]
\centering
\includegraphics[width=.95\textwidth]{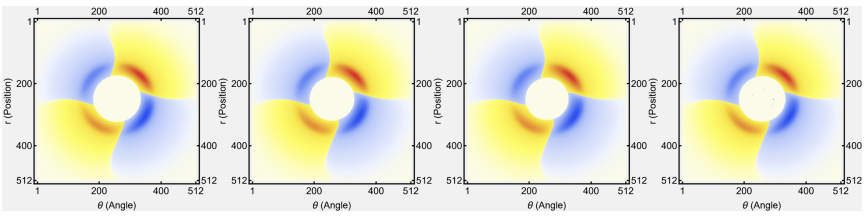}
\includegraphics[width=.95\textwidth]{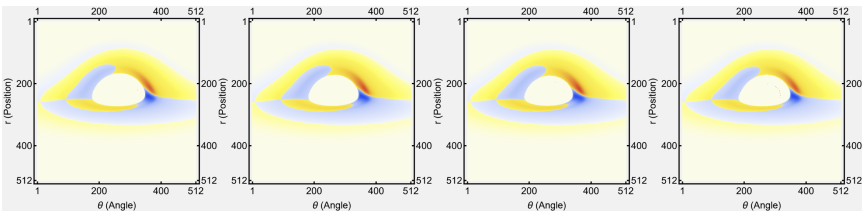}
\caption{Same as Fig.~\ref{fig:7}, but for the retrograde accretion flow.}
\label{fig:8}
\end{figure}

Figures~\ref{fig:7} and \ref{fig:8} show the complementary Stokes-$U$ maps. At low inclination, $U$ displays a more continuous twisted or rotational pattern around the shadow than does $Q$. At $\theta_0=75^\circ$, the pattern is projected into curved bands and layered structures above and below the disk. This difference does not imply a separate propagation mechanism: $Q$ and $U$ are two orthogonal projections of the same linear-polarization field, controlled by $\cos2\chi$ and $\sin2\chi$, respectively. A continuous EVPA field can therefore produce markedly different two-dimensional morphologies in the two Stokes bases.

Changing $q$ at fixed $a=0.5$ preserves the overall twisted or banded pattern but modifies its local curvature, enhancement locations, and width. The changes are not described by a simple monotonic increase or decrease of a single Stokes amplitude. Instead, the tidal charge reorganizes both the photon mapping and the polarization orientation, and the trigonometric response of $U$ converts the same EVPA changes into a morphology distinct from that of $Q$. Increasing the spin produces further local changes without removing the characteristic large-scale structure. Taken together, $Q$ and $U$ therefore provide complementary information on polarization direction that is not available from $I_{\rm pol}$ alone.

\subsection{$Q$--$U$ polarization-state space}

\begin{figure}[htbp]
\centering
\includegraphics[width=.95\textwidth]{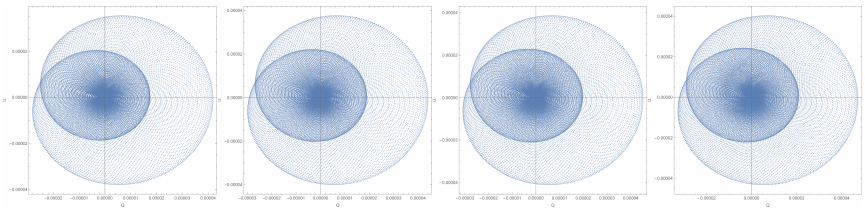}\\[2mm]
\includegraphics[width=.95\textwidth]{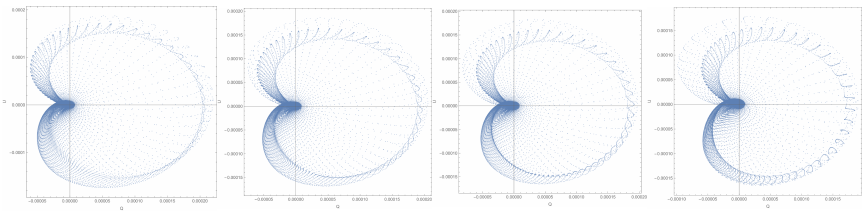}
\caption{$Q$--$U$ plane distributions for the prograde accretion flow around a rotating braneworld black hole with $M=1$. In each panel, the horizontal axis represents Stokes $Q$ and the vertical axis represents Stokes $U$. The upper and lower rows correspond to $\theta_0=17^\circ$ and $75^\circ$, respectively. From left to right, the parameter combinations are $(a,q)=(0.5,-0.3)$, $(0.5,0.1)$, $(0.5,0.3)$, and $(0.8,0.3)$.}
\label{fig:9}
\end{figure}

\begin{figure}[htbp]
\centering
\includegraphics[width=.95\textwidth]{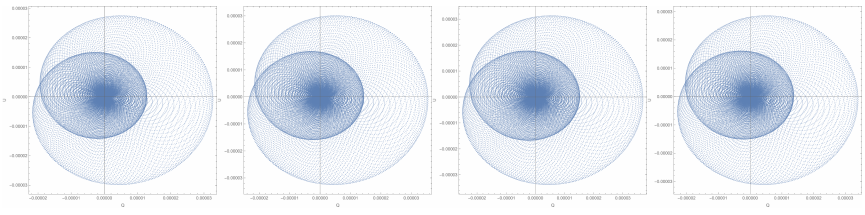}\\[2mm]
\includegraphics[width=.95\textwidth]{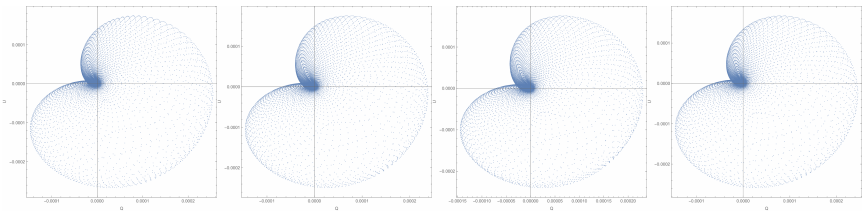}
\caption{Same as Fig.~\ref{fig:9}, but for the retrograde accretion flow. In each panel, the horizontal axis is Stokes $Q$ and the vertical axis is Stokes $U$.}
\label{fig:10}
\end{figure}

The $Q$--$U$ plane combines the polarization amplitude and orientation in a single state space. The radial coordinate in this plane is $\sqrt{Q^{2}+U^{2}}=I_{\rm pol}$, while the polar angle is $2\chi$. Figures~\ref{fig:9} and \ref{fig:10} therefore show whether changes in $q$, $a$, inclination, or orbital sense reorganize the full linear-polarization state even when the corresponding intensity maps remain similar.

At $\theta_0=17^\circ$, all parameter sets show comparatively regular multilayer ringlike or elliptical distributions. Changing $q$ preserves this overall state-space type but shifts the dense central region, outer scale, and local curvature. These changes combine a radial contribution from the modified polarized intensity with an angular contribution from the transported EVPA. At $\theta_0=75^\circ$, the distributions become strongly stretched toward one side and develop more pronounced bending and local curling. The enhanced deformation reflects both the stronger Doppler asymmetry and the greater asymmetry among strong-field propagation histories at high inclination.

The prograde and retrograde models have broadly similar $Q$--$U$ organization at low inclination, whereas their extension direction and local curvature differ more visibly at $75^\circ$. These differences arise from the combined changes in Doppler/redshift mapping, photon initial conditions, and source-side polarization set by the emitter four-velocity. The key result is therefore not the appearance of a wholly new polarization-state morphology, but a systematic reorganization of an existing one.

\begin{figure}[htbp]
\centering
\includegraphics[width=.95\textwidth]{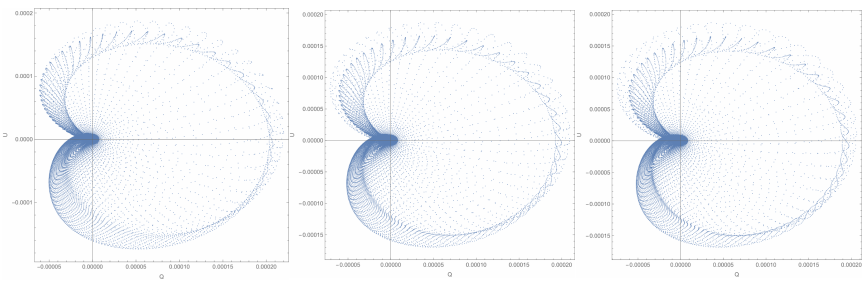}
\caption{Comparison of the $Q$--$U$ plane distributions for the prograde accretion flow around a braneworld black hole, a Kerr black hole, and a Kerr--Newman black hole, with $M=1$ and $\theta_0=75^\circ$. In each panel, the horizontal axis denotes Stokes $Q$ and the vertical axis denotes Stokes $U$. From left to right, the panels correspond to the braneworld model $(a,q)=(0.5,-0.3)$, the Kerr model with $a=0.5$, and the Kerr--Newman model $(a,Q)=(0.5,0.3)$.}
\label{fig:11}
\end{figure}

Figure~\ref{fig:11} compares the braneworld model with Kerr and Kerr--Newman at fixed $a=0.5$, $\theta_0=75^\circ$, and prograde orbital motion. The three spacetimes share the same broad asymmetric, curved $Q$--$U$ morphology, but clear local differences remain. The most visible difference occurs in the upper-left region, $Q<0$ and $U>0$: the braneworld trajectories are more tightly packed and the outer envelope is more compact, the Kerr distribution is broader, and the Kerr--Newman outer trajectories spread farther outward. By contrast, the large-scale envelope in the $U<0$ region is more similar across the three cases, with differences concentrated mainly in local curvature and outer-edge position. The comparison therefore indicates a nonuniform reorganization of polarization-state space rather than a simple global expansion or contraction.

The local tightening or broadening of the trajectories encodes a joint change in polarized intensity and EVPA. In the braneworld model the more concentrated upper-left trajectories occupy a narrower region of state space, whereas the broader Kerr and Kerr--Newman envelopes correspond to a wider range of polarization-amplitude and orientation combinations. Importantly, this sequence should not be interpreted as a continuous variation of a single charge parameter. The Kerr--Newman parameter $Q$ is a four-dimensional Maxwell charge, whereas the braneworld tidal charge $q$ is an effective projection of higher-dimensional bulk gravity onto the brane and can be negative. The geometry associated with $q=-0.3$ therefore has no direct realization as an ordinary four-dimensional Kerr--Newman electric charge.

This comparison suggests that an extra-dimensional gravitational imprint need not manifest itself through a qualitatively new macroscopic polarization image. It can instead appear as local compression, expansion, and concentration of trajectories in specific regions of polarization-state space. Quantitative measures of local envelope width and occupied $Q$--$U$ area may therefore provide useful diagnostics in future attempts to identify higher-dimensional geometric effects in horizon-scale polarimetry.

\subsection{Cumulative net linear polarization degree}

For a circular aperture extending from the inner emitting region to radius $r$, we define cumulative Stokes quantities by
\begin{align}
 I_{\rm cum}(r)&=\int_{\mathcal A(r)} I\,d\Omega,\notag\\
 Q_{\rm cum}(r)&=\int_{\mathcal A(r)} Q\,d\Omega,\notag\\
 U_{\rm cum}(r)&=\int_{\mathcal A(r)} U\,d\Omega.
\label{eq:cumulative_stokes}
\end{align}
and the corresponding cumulative net linear polarization degree as
\begin{equation}
 \Pi_{\rm net}(r)=\frac{\sqrt{Q_{\rm cum}^{2}(r)+U_{\rm cum}^{2}(r)}}{I_{\rm cum}(r)}.
\label{eq:Pi_net}
\end{equation}

\begin{figure}[htbp]
\centering
\includegraphics[width=.98\textwidth]{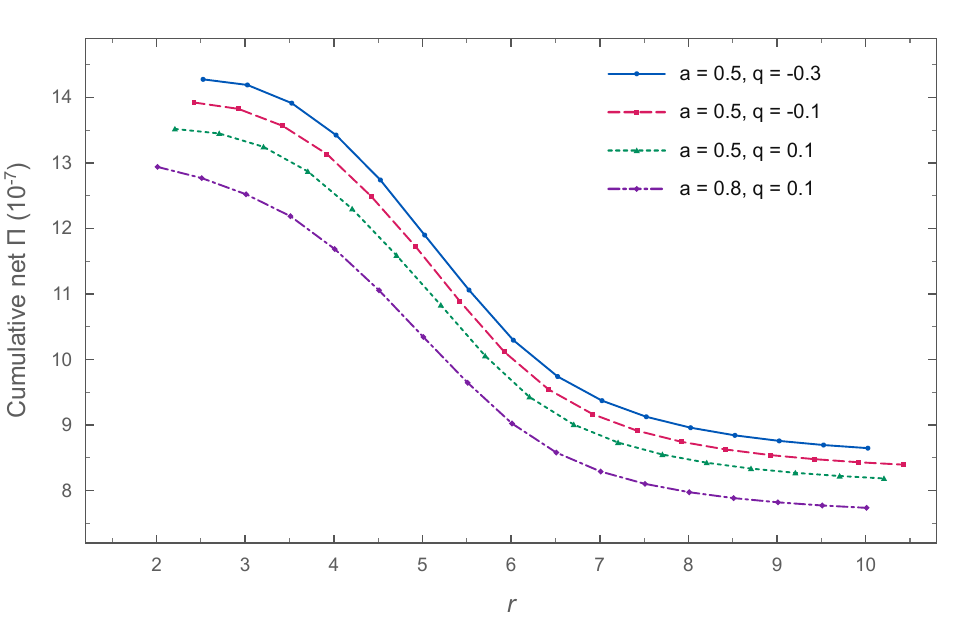}
\caption{Cumulative net linear polarization degree $\Pi_{\rm net}$ as a function of radial coordinate $r$ for different tidal charges and spins. The blue solid, red dashed, green dash-dotted, and purple dash-dotted curves correspond to $(a,q)=(0.5,-0.3)$, $(0.5,-0.1)$, $(0.5,0.1)$, and $(0.8,0.1)$, respectively.}
\label{fig:12}
\end{figure}

Figure~\ref{fig:12} shows $\Pi_{\rm net}(r)$ for four representative parameter combinations. All curves decrease as the cumulative radius grows and approach a flatter dependence at large $r$. The decrease is most pronounced over approximately $r\simeq4$--$6.5$, while the variation becomes weaker for $r\gtrsim7$. As the aperture expands, additional emission increases the total intensity, but polarization vectors from different spatial regions need not be aligned. Partial cancellation between the integrated $Q$ and $U$ contributions therefore reduces the net polarization fraction.

At fixed $a=0.5$, the ordering is $q=-0.3$, $-0.1$, and $0.1$ from highest to lowest $\Pi_{\rm net}$. Thus increasing the tidal charge across these representative values slightly lowers the cumulative net polarization degree without changing the common radial trend. This result can be interpreted as a change in the amount of spatial depolarization produced by the $q$-dependent lens mapping and polarization transport, rather than as a direct measurement of a separate electromagnetic effect. At fixed $q=0.1$, increasing the spin from $a=0.5$ to $0.8$ further lowers $\Pi_{\rm net}$, with the separation becoming more visible at intermediate and larger radii. The spin-dependent rotational geometry changes the azimuthal photon propagation and polarization transport and thereby modifies the degree of vector cancellation after spatial integration.

The cumulative polarization is controlled jointly by the local polarized emissivity, null geodesics, redshift, lensing, and parallel transport. The common shape of the four curves indicates that the radial emission profile and aperture depolarization set the dominant trend, while their stable amplitude offsets encode the additional modulation by $q$ and $a$. The parameter dependence reported here should therefore be understood within the fixed emission and magnetic-field model adopted in this work.

\section{Conclusions}
\label{sec:conclusion}

We have studied synchrotron polarization transport and polarized imaging of an optically thin accretion flow around a rotating braneworld black hole. By directly integrating the photon null-geodesic equations together with the parallel transport of the polarization four-vector, we obtain the photon trajectories and polarization states without relying on a Walker--Penrose constant. The same local synchrotron emissivity and magnetic-field prescription is used throughout the parameter study, allowing the geometric roles of the tidal charge $q$, black-hole spin $a$, observer inclination $\theta_0$, and prograde or retrograde flow kinematics to be compared on a common footing.

Within the parameter range explored here, neither $q$ nor $a$ produces a qualitatively new large-scale polarization morphology. Their main signatures are local changes in polarized intensity and orientation in the strong-field region near the shadow boundary and bright ring. The viewing inclination has the largest impact on the global projected morphology: low-inclination images remain close to ringlike, whereas high-inclination images become strongly compressed and asymmetric as projection and relativistic Doppler effects are enhanced. Spin and tidal charge then reorganize the local structure through their effects on null geodesics, redshift, lensing, and polarization transport, with frame dragging providing an additional spin-dependent modification of azimuthal propagation.

The direction-sensitive observables expose differences that are less evident in $I_{\rm pol}$. Stokes $Q$ highlights spatial partitioning and local boundaries, while $U$ emphasizes the continuous twisted and banded structures associated with the same EVPA field in an orthogonal Stokes projection. Their joint $Q$--$U$ representation combines polarization amplitude and orientation in a single state space. Low-inclination models form relatively regular multilayer ringlike or elliptical distributions, whereas high-inclination models show stronger asymmetric stretching, bending, and local curling. The prograde--retrograde distinction is also more visible at high inclination because the emitter kinematics alter the Doppler/redshift mapping and source-side polarization conditions.

A direct comparison among braneworld, Kerr, and Kerr--Newman spacetimes shows similar global $Q$--$U$ envelopes but systematic local differences in trajectory concentration, outer extent, and curvature. In particular, the braneworld configuration considered here is more compact in the $Q<0$, $U>0$ region than the Kerr and Kerr--Newman reference cases. This distinction is physically relevant because the braneworld tidal charge is not a Maxwell charge: it is an effective imprint of higher-dimensional bulk gravity on the four-dimensional brane and can take negative values. The observed local state-space reorganization therefore provides a concrete example of how an extra-dimensional geometric modification can enter a four-dimensional polarization observable without generating an entirely new macroscopic image.

The cumulative net linear polarization degree provides a complementary integrated diagnostic. For all representative parameter combinations it decreases with increasing cumulative radius and then gradually flattens, consistent with increasing vector cancellation among Stokes contributions from different regions. At fixed $a=0.5$, increasing $q$ from negative to positive values lowers $\Pi_{\rm net}$, while at fixed $q=0.1$, increasing the spin from $a=0.5$ to $0.8$ produces a further reduction. Because the absolute level of $\Pi_{\rm net}$ depends on the adopted source and magnetic-field model, we emphasize the relative differences among parameter sets rather than a model-independent absolute prediction.

Taken together, $I_{\rm pol}$, EVPA, Stokes $Q$ and $U$, the $Q$--$U$ state space, and the cumulative net polarization degree probe different aspects of the same transport problem. Their joint use provides complementary constraints on background geometry, viewing inclination, and accretion-flow kinematics, and offers a potential route toward constraining braneworld tidal charge with future horizon-scale polarimetric observations.

\section*{Acknowledgements}

This work was supported by the Sichuan Provincial Key Projects for Science-Education Integration 
(Grant No.~25LHJJ0097).

\end{document}